\documentstyle[aps,prd,amssymb,12pt,preprint,epsf]{revtex}
\tightenlines
\newcommand{\be}{\begin{equation}}
\newcommand{\ee}{\end{equation}}

\begin{document}
\draft
\title{
Equation of state and transport processes in
 self--similar spheres}
\author{W. Barreto${}^{1,2}$,
	C. Peralta${}^{1}$ and
	L. Rosales${}^{3}$}
\address{
${}^{1}$Laboratorio de F\'{\i}sica Te\'orica,
         Departamento de F\'{\i}sica,
         Escuela de Ciencias, N\'ucleo de Sucre,
         Universidad de Oriente,
         Cuman\'a, Venezuela\\
${}^{2}$Centro de Astrof\'{\i}sica
         Te\'orica,
         Universidad de los Andes,
         M\'erida, Venezuela\\
${}^{3}$ Secci\'on de F\'{\i}sica,
	 UNEXPO Antonio Jos\'e de Sucre, Puerto Ordaz, Venezuela}
\maketitle
\begin{abstract}
We study the effect of transport processes
(diffusion and free--streaming)
on a collapsing spherically symmetric distribution of matter in a
self--similar space--time. 
A very simple solution shows 
interesting
features when it is matched with the Vaidya exterior solution. In the mixed
case (diffusion and free--streaming), we find a barotropic
equation of state 
in the stationary regime. In the diffusion approximation the 
gravitational potential at the surface is always constant; if
we perturb the stationary state, the system is very
stable, recovering the barotropic equation of state as time
progresses. 
In the free--streaming case the self--similar evolution is stationary but
with a non--barotropic equation of state. 
\end{abstract}
\section{Introduction}
The formation of compact objects is usually preceded by an epoch of
 radiative collapse
\cite{fjlis92}. One of the main difficulties in the study of these systems is
that there is no reliable information about the equation of
state in the central regions of superdense stars, such as neutron stars,
 and this
leads to assumptions of a very general nature \cite{t90}. 

Few exact solutions to the Einstein equations are relevant to gravitational
collapse. 
For this reason, new collapse solutions are very useful, even if they
are simplified ones \cite{op90}.
It is well known that the field equations admit homothetic motion 
\cite{op90,ct71,hp91,lz90}. 
Applications of homothetic simila\-rity range from modeling 
black holes to producing counterexamples to the cosmic censorship 
conjecture \cite{ch74,bh78a,bh78b,cy90,eimm86,l92,b95,cc98}.  

The natural formulation of the collapse problem is that of initial values, 
with an {\it a priori} defined equation of state. Ho\-wever, there are only 
a few 
solutions to the Einstein field equations with a well defined equation of 
state. It has been shown that the only perfect fluid equation of state 
compatible with self-similarity is the barotropic one \cite{ct71}.

Emission of photons or neutrinos is a typical process in the evolution
of massive stars.
The only plausible mechanism to deliver almost all the
binding gravitational energy, during the collapse toward a neutron
star, is that of neutrino emission \cite{h96}. It seems clear that the 
free--streaming process is associated with the initial stages of 
the collapse, while 
the diffusion approximation becomes valid toward the final stages.

In this paper, we explore self--similar gravitational co\-llapse. 
For such an assumption, we recast the geometrical variables so as to have an 
explicit radial dependence. The fluid is considered with heat flow 
(the diffusion approximation) or free propagation in the radial direction
(free--streaming).  The interior solution is matched to an exterior
(Vaidya) solution by means of the Darmois--Lichnerowicz conditions.

We discuss neither the microprocesses that produce free--streaming, nor the
temperature distribution during diffusion. For the latter another
approach is necessary to avoid pathological behavior \cite{m96}.

\section{The field equations and matching}
For the geometric description of the interior of the distribution we use
the radiation metric in the spherical Bondi form \cite{b64}
\begin{equation}
\label{mdb}ds^2=e^{2\beta }\left(\frac{V}{r}\, du^2+du\,dr\right)
-r^2\Biggl( d\theta ^2+\sin^2 \theta\, d\phi ^2\Biggr) 
\end{equation}
where $\beta $ and $V$ are functions of $u$ and $r$. Here $u$ is a time--like
coordinate, $r$ is a null coordinate ($g_{rr}=0$), that is,
$r\ge 0$ is an affine parameter along the null generators of $u=$ constant null
hypersurfaces, and $\theta$, 
$\phi$ are the usual angular coordinates; we are using geometrized units
($c=G=1$).

The hydrodynamic scenario, as viewed by a local Minkowskian observer comoving
with the fluid (with velocity $-\omega$), consists of an isotropic fluid of 
density $\rho$ and pressure
$p$, unpolarized energy density $\epsilon$ and heat flux $q$ 
traveling in the radial direction.
Therefore, for this comoving observer, the covariant energy--momentum tensor
is
$$
\left( 
\begin{array}{cccc}
\rho + \epsilon & -q -\epsilon & 0 & 0 \\ 
-q -\epsilon & p + \epsilon & 0 & 0 \\ 
0 & 0 & p & 0 \\ 
0 & 0 & 0 & p
\end{array}
\right )_{.} 
$$
Note that the velocity of matter in the Bondi coordinates is
\begin{equation}
\frac{dr}{du}=\frac{V}{r}\frac{\omega}{1-\omega}\,. \label{eq:mv}
\end{equation}
We can write the Einstein field equations 
as \cite{bn91}
\begin{equation}
\label{eq:ec1}
\frac{4\pi r^2}{(1-\omega^2)}[\rho + p \omega^2 + 2 \omega q+
\epsilon (1+\omega)^2]
= \tilde m_{,r}-\frac{ e^{-2\beta} \tilde m_{,u} }{ (1 - 2 \tilde m/r) } ,
\end{equation}

\begin{equation}
\label{ec2}\frac{4\pi r^2}{(1+\omega)}[(\rho-\omega p) - (1-\omega)q]
= \tilde m_{,r} ,
\end{equation}

\begin{equation}
\label{ec3} \frac{ 2\pi r (1-\omega)(\rho+p -2q) }{ (1-2\tilde m/r)(1+\omega) }
= \beta_{,r} ,
\end{equation}
\begin{equation}
\label{ec4}
8\pi p = -2e^{-2\beta}\beta_{,ur}+[3\beta_{,r} (1-2\tilde m_{,r})
-\tilde m_{,rr}]/r +
 (1 -2\tilde m/r)
(2\beta_{,rr}+4\beta^2_{,r}-\beta_{,r}/r) ,
\end{equation}
where the comma subscript represents partial differentiation with respect to the
indicated coordinate and $\tilde m$ is the Bondi mass defined by 
$\tilde m =(r-V \exp(-2\beta))/2$.

The exterior space--time is described by the Vaidya radiating metric
\cite{v51}. In order to match this to the interior solution, 
we use the 
Darmois--Lichnerowicz conditions. These 
are equivalent to the continuity of the functions $\beta $ and $\tilde m$ 
across the boundary of the sphere, and to the condition \cite{hj83,hjr80}
\begin{equation}
[-\beta_{,u}e^{2\beta} + (1-2\tilde m/r)\beta_{,r} - 
\tilde m_{,r}/(2r)]_{a}=0\, ,
\label{eq:sfc}
\end{equation}
which is equivalent to $p_a=q_a$. The subscript $a$ indicates that the quantity is being evaluated at the
surface $r=a(u)$.
\section{Spherical and self--similar interior solutions}
Self--similarity is invariably defined by the existence of a homothetic
Killing vector field \cite{ct71}. 
A homothetic vector field on the manifold is one that satisfies 
$\pounds_\xi {\bf g}=$2$n{\bf g}$ on a local chart,
 where $n$ is a constant on the 
manifold and $\pounds$ denotes the Lie derivative operator. If $n \ne 0$ we
have a proper homothetic vector field and it can always be scaled to have
$n = 1$; if $n = 0$ then $\xi$ is a Killing vector on the manifold 
\cite{h88,h90,c94}. So, for a constant rescaling, $\xi$ satisfies $
\pounds_\xi{\bf g}=$2${\bf g}$ and has the form 
$
\xi =\Lambda (u,r)\partial_u  +\lambda (u,r)\partial_r
$. 
If the matter field is a perfect fluid, the only
equation of state consistent with $\pounds_\xi{\bf g}=$2${\bf g}$
is a barotropic one \cite{ct71}. The homo\-thetic
 equations reduce to $\xi(X)=0$, $\xi(Y)=0$, $\lambda=r$ and
$\Lambda=\Lambda(u)$, where $X\equiv\tilde m/r$ and $Y\equiv\Lambda e^{2\beta}/r$.
Therefore, $X=X(\zeta)$ and $Y=Y(\zeta)$ are solutions if the self--similar
 variable is defined as $\zeta\equiv r\, \exp(- \int du/\Lambda)$. 
 Here we assume that
 $X=C_1\zeta^k$ and that $Y=C_2\zeta^l$, where $C_1$, $C_2$, $k$ and $l$ are
constants. This power--law dependence on $\zeta$ is based on
the fact that any  function of
$\zeta$ is solution of $\pounds_\xi {\bf g}=$2${\bf g}$. As we shall see, this
simplifying assumption is not devoid of physical meaning.
Demanding continuity of the first fundamental form we get
the following metric solutions:
\be
\label{mtilde} \tilde{m}= \tilde m_a(r/a)^{k+1},
\ee
\be
\label{e2b} e^{2\beta} = (r/a)^{l+1}.
\ee
Condition (\ref{eq:sfc}) 
then implies that the local radial velocity $\omega$ is determined at
the surface in terms of the gravitational potential $\tilde m_a/a$
and the parameters
$k$ and $l$:
\begin{equation}
\omega_a=1-\frac{(1-2\tilde m_a/a)}{\tilde m_a/a}\frac{(1+l)}{(1+k)}\,\, .
\label{eq:omegaa}
\end{equation} 
Equations (\ref{eq:mv}) and (\ref{eq:ec1}) evaluated at
the surface constitute the system of (ordinary differential) equations
 in $a(u)$ and $\tilde m_a(u)$, to be integrated while taking into 
account Eq. (\ref{eq:omegaa}).
 Thus, the dynamics at the surface is completely determined if we 
establish how energy is exchanged with the exterior. In fact, we have found that
self--similarity determines the luminosity profiles \cite{bc95,bor98}.

\section{Transport processes}
In order to explore the effect of the transport processes on 
self--similar gravitational collapse, 
we consider below a combination
of the diffusion and free--streaming mechanisms, and the action of each one
separately. In particular we discuss whether the barotropic equation of state
holds in each case. 

\subsection{Mixed}
If the transport mechanism is mixed, we require additional information.
Only in this case do we suppose
orthogonality between the four--velocity and the homothetic vector.
This condition has been employed to obtain static solutions \cite{autobus}
and it also establishes a relationship between the homothetic vector and the
 equation of state \cite{kg96}.
Another meaning attributable to the orthogonality condition is that the group,
generated by the homothetic vector, is acting upon the three--space
 comoving with the
observer. The referred condition, together with
$\pounds_\xi {\bf g}=$2${\bf g}$,
can thus be seen as a covariant definition of self--similarity \cite{h98}.
Therefore, from the orthogonality condition 
we obtain the temporal component of the homothetic vector:
\begin{equation}
 \Lambda=\tilde m_a (1+l)/(1+k). \label{eq:lambda}
\end{equation}
The radiation flux
at the surface is then determined from the symmetry equations.
The heat flux at the surface is also determined from
equation (\ref{ec4}) or, equivalently, from $T^{\mu}_{1;\mu}=0$
evaluated at the surface. 
Now, feeding back (\ref{eq:omegaa})
and (\ref{eq:lambda}) into the symmetry equations,
we find algebraically by means of symbolic manipulation with REDUCE,
two restrictions which allow us to satisfy the symmetry equations:
(i) $k=-l$; (ii) $k=[(1-2\tilde m_a/a)(l+2)-1]/(2\tilde m_a/a)$.
Restriction (ii) leads to $\omega_a=-1$ (through Eq. (\ref{eq:omegaa})),
 that is, the fluid collapses
at light speed; we must reject this possibility. 
Under restriction (i), $k=-l$,
the homothetic symmetry is preserved at all points of the 
space--time, when we integrate numerically (using the Runge--Kutta method)
the system of
equations at the surface.  Specifically, for $k=0$ the physical variables
are well behaved and the equation of state is barotropic
in the whole sphere; that is, the ratio $p/\rho$ is constant
at all points of the material (see Figure 1). 
All shells collapse with the emission of energy. The pressure, 
density, heat flow and radiation flux are
stationary inside the sphere. The matter velocity decreases toward
the center and is constant at the surface. Observe that $p\approx(3/5)\rho$,
as we expect, because the heat flow diminish the effective gravitation or,
in others words, the equation of state is softened by the diffusive
process \cite{bhs89,bn91}.
For $k\ne 0$ the equation of
state is not longer barotropic and the physical variables are not stationary,
although the interior space--time is
self--similar.

\subsection{Diffusion}
In the diffusion approximation ($\epsilon=0$) we again obtain the heat flux
at the surface from equation (\ref{ec4}) or $T^{\mu}_{1;\mu}=0$
evaluated at the surface. From the symmetry equations evaluated
at the surface, we deduce that the gravitational potential at the
surface, $\tilde m_a/a$, is a function of only $k$ and $l$.
The surface equations thus reduce to one differential equation for $a$
(or $\tilde m_a$). This situation leads us to four possible restrictions:
(i) $k=-l$; (ii) $l=-1$ which is equivalent to $\omega_a=1$ 
(see Eq. (\ref{eq:omegaa})), that is,
the fluid explodes at light speed; 
(iii) a complicated polynomial of degree three in $k$
with coefficients depending on $l$;
(iv) a complicated polynomial of degree seven in $k$ with coefficients
depending on $l$. 
We solved analitycally the polynomial of degree three
(using REDUCE), obtaining one real and two complex solutions.
 All these
must be rejected because they do not have physical meaning for a 
wide interval of numerical values of $l$.  After solving numerically
the polynomial 
of degree seven, we found that the  physically acceptable models 
are similar to those that emerge from restriction (i), $k=-l$.
In any case, the gravitational potential at the surface is
 constant ($\tilde m_a
/a$ depending only on $k$)
although the distribution collapses. If we perturb 
this stationary state, we find
numerically that the system recovers
the barotropic equation of state only for $k=0$ (see Figure 2). 
Perturbations consist of enhancing or diminishing the gravitational
potential at the surface, while preserving the values of $k$. Therefore, we must
integrate numerically (using the Runge--Kutta method again) the two perturbed 
differential equations at the surface. It is important to stress that the
symmetry equations are satisfied everywhere and at all times when we
perturb the system.

\subsection{Free--streaming}
In this transport process ($q=0$) the boundary condition 
reduces to $p_a=0$. We obtain again
a gravitational potential which is constant at the surface
 (from $T^{\mu}_{1;\mu}$
evaluated at the surface). Therefore, the luminosity profile 
and the temporal component of the homothetic vector are determined
from one of the symmetry equations. No restrictions appear on the
parameters $k$ and $l$, and no barotropic behavior is found, at least for
the cases examined.

\section{Conclusion}
We have studied the effect
of diffusion and free--streaming  on self--similar gravitational collapse
using the solutions to the Einstein equations given by (\ref{mtilde})
 and (\ref{e2b}).
We find that the equation of state is barotropic only
when diffusion occurs and the parameter
$k$ is zero. In general ($k\ne 0$) the equation of state is not barotropic,
although self--similarity holds everywhere.
Thus for 
non--perfect fluids self--similarity is connected with
a barotropic equation of state only under special conditions. The heat flow
seems to be crucial for this result. Viscosity in self--similar
 distributions has also been considered previously and the same result emerges;
that is, the equation of state can be barotropic \cite{bor98}. 
We have found relativistic and self--similar examples 
compatible with a nonbarotropic equation of state 
\cite{pz97},  at least within
our chosen simple solutions and the orthogonality condition between
the generator of the group and the four--velocity. It would be of
interest to attempt to relax some of our suppositions in order to
investigate how general our conclusions are.

\acknowledgements
We benefited from research support by the Consejo de Investigaci\'on
under Grant CI-5-1001-0774/96 of the Universidad de Oriente. 
We would like to thank Loren Lockwood for his helpful reading and 
valuable comments.

\begin{figure}
\centerline{\epsfxsize=5.5in\epsfbox{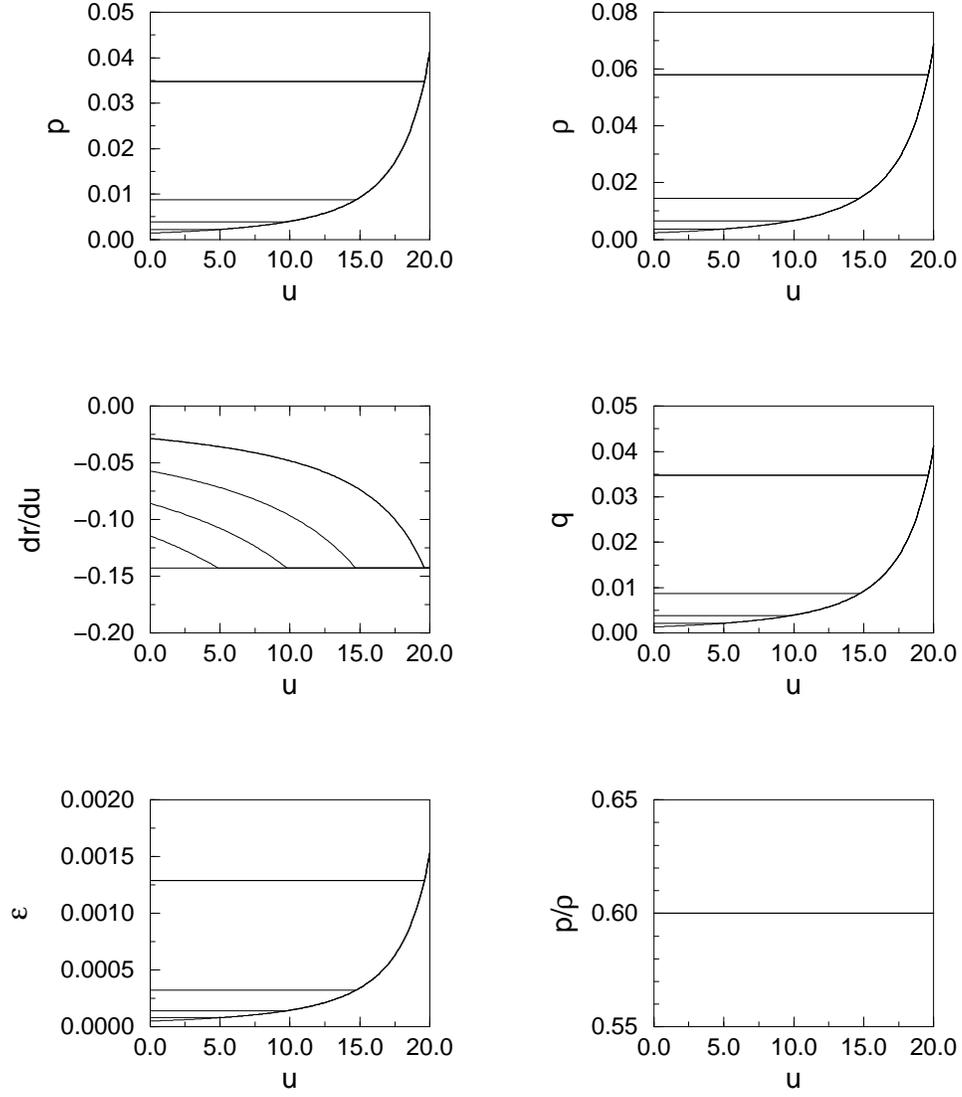}}
\caption{Evolution of the physical variables for the mixed 
transport process 
with $k=0$. In each graph the curves correspond (from the uppermost
to the lowermost) to ratios: $r/a(0)=0.2$, $0.4$, $0.6$, $0.8$ and $1.0$,
 respectively. The ratio $p/\rho$ is the same at any point of the material.}
\label{fig:pv}
\end{figure}

\begin{figure}
\centerline{\epsfxsize=6in\epsfbox{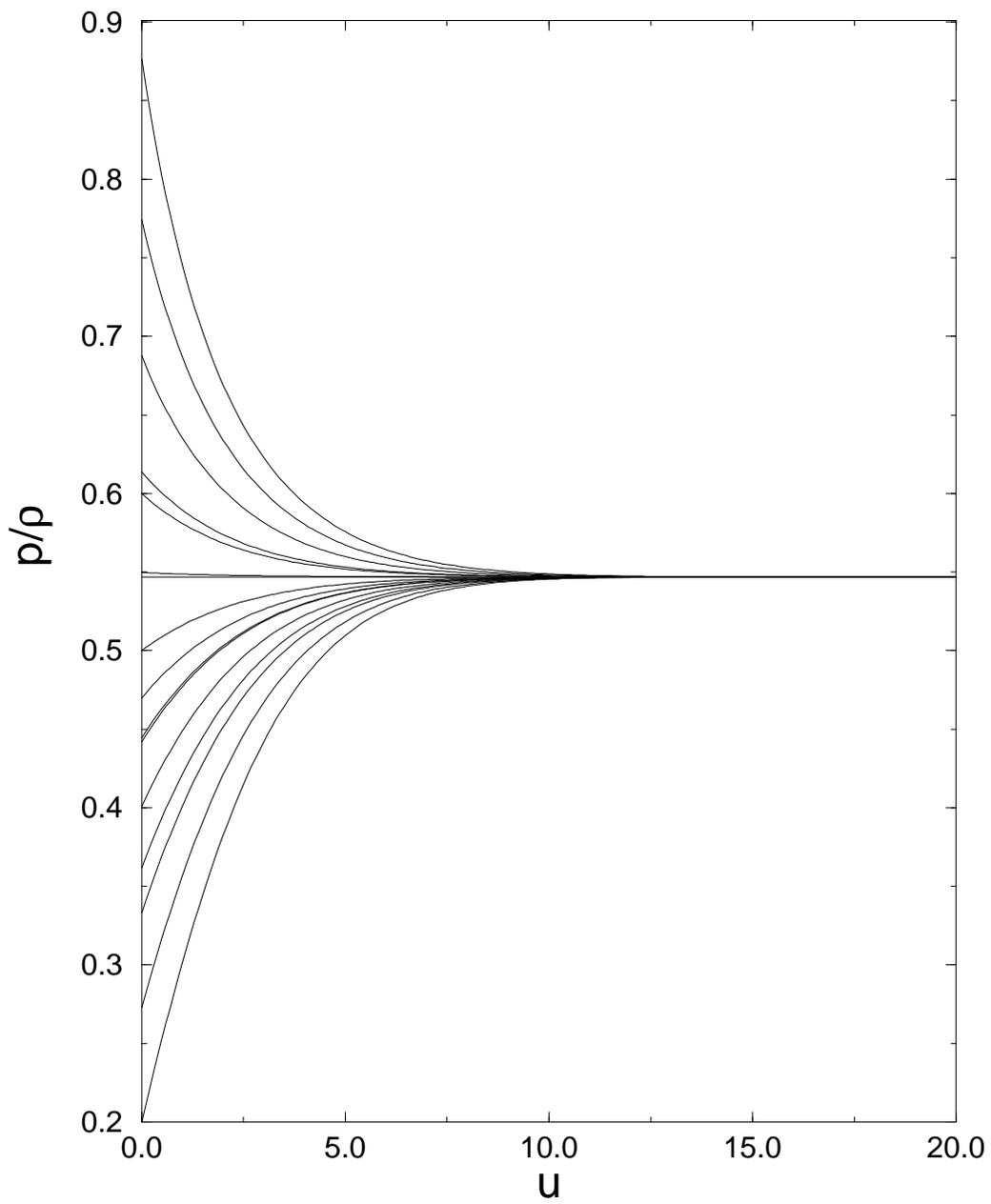}}
\caption{The ratio $p/\rho$ as a function of retarded time
 in the diffusion approximation
with $k=0$, considering perturbations of the stationary regime. }
\label{fig:poverho}

\end{figure}


\begin{thebibliography}{99}

\bibitem{fjlis92} F. Fayos, X. Ja\'en, E. Llanta and J. Senovilla,
 Phys. Rev. D {\bf 45}, 2732 (1992).

\bibitem{t90} R. Tikekar, J. Math. Phys. {\bf 31}, 2454 (1990). 

\bibitem{op90} A. Ori and T. Piran, Phys. Rev. D {\bf 42}, 1068 (1990).

\bibitem{ct71}  M. E. Cahill and A. H. Taub, Commun. Math. Phys.
{\bf 21}, 1 (1971).

\bibitem{hp91} R. N. Henriksen and K. Patel, Gen. Rel. Grav.
{\bf 23}, 527 (1991).

\bibitem{lz90} K. Lake and T. Zannias, Phys. Rev. D
{\bf 41}, 3866 (1990).

\bibitem{ch74} J. Carr and S. Hawking, MNRAS {\bf 168}, 399 (1974).

\bibitem{bh78a} G. V. Bicknell and R. N. Henriksen, 
Ap. J. {\bf 219}, 1043 (1978).

\bibitem{bh78b} G. V. Bicknell and R. N. Henriksen,
Ap. J. {\bf 225}, 237 (1978).

\bibitem{cy90} J. Carr and A. Yahil, Ap. J.  {\bf 360}, 330 (1990).

\bibitem{eimm86} D. M. Eardley, J. Isenberg, J. Marsden and V. Moncrief,
 Comm. Math. Phys. {\bf 106}, 137 (1986).
 
\bibitem{l92} K. Lake, Phys. Rev. Lett. {\bf 68}, 3129 (1992).

\bibitem{b95} P. R. Brady, Phys. Rev. D {\bf 51}, 4168 (1995).

\bibitem{cc98} B. J. Carr and A. A. Coley, gr--qc/9806048.

\bibitem{h96}  L. Herrera, {\it Campos gravitacionales en la materia:
la otra cara de la moneda}. ``II Escuela Venezolana de Relatividad,
Campos y Astrof\'{\i}sica''. Facultad de Ciencias, 
Universidad de los Andes, M\'erida, Venezuela (1996).

\bibitem{m96} J. Mart\'\i nez,  Phys. Rev. D {\bf 53}, 6921 (1996).

\bibitem{b64} H. Bondi, Proc. Roy.
Soc. London A {\bf 281}, 39 (1964).

\bibitem{bn91} W. Barreto and  L. N\'u\~nez, Ap. Sp. Sc. {\bf 178}, 
              261 (1991).

\bibitem{v51} P. C. Vaydia, Phys. Rev. {\bf 83}, 10 (1951).

\bibitem{hj83} L. Herrera and J. Jim\'enez, Phys. Rev. D {\bf 28}, 
2987 (1983).

\bibitem{hjr80} L. Herrera, J. Jim\'enez, and G. Ruggeri, Phys. Rev. D
{\bf 22}, 2305 (1980).

\bibitem{h88} G. S. Hall, Gen. Rel. Grav. {\bf 20}, 671 (1988).

\bibitem {h90} G. S. Hall, J. Math. Phys. {\bf 31}, 1198 (1990).

\bibitem {c94} J. Carot, L. Mas and A. M. Sintes, J. Math. Phys.
{\bf 35}, 3560 (1994).

\bibitem{bc95} W. Barreto and L. Castillo, J. Math. Phys. {\bf 36}, 5789 (1995).

\bibitem{bor98} W. Barreto, J.  Ovalle and B. Rodr\'\i guez,
 Gen.  Rel. Grav. {\bf 30}, 15 (1998). 

\bibitem{autobus} L. Herrera, J. Jim\'enez, L. Leal, J. Ponce de Le\'on,
 M. Esculpi and V. Galina, J. Math. Phys. {\bf 25}, 3274 (1984).

\bibitem{kg96} C. Kolassis and J. B. Griffiths, Gen. Rel. Grav. {\bf 28},
 805 (1996).

\bibitem{h98} L. Herrera (private communication).

\bibitem{bhs89} W. Barreto, L. Herrera and N. Santos, Ap. J. {\bf 344},
 158 (1989).

\bibitem{pz97} D. Pollney and T. Zannias, Phys. Rev. D {\bf 56}, 8086 (1997).

\end{thebibliography}
\end{document}